\documentclass{PoS}
\usepackage{amsmath,amssymb,amsfonts,color,graphicx,cite,soul}
\usepackage{subfigure}

\newcommand{\bq}{\begin{align}}
\newcommand{\eq}{\end{align}}

\catcode`@=11
\def\citer{\@ifnextchar [{\@tempswatrue\@citexr}{\@tempswafalse\@citexr[]}}
 
\def\@citexr[#1]#2{\if@filesw\immediate\write\@auxout{\string\citation{#2}}\fi
  \def\@citea{}\@cite{\@for\@citeb:=#2\do
    {\@citea\def\@citea{--\penalty\@m}\@ifundefined
       {b@\@citeb}{{\bf ?}\@warning
       {Citation `\@citeb' on page \thepage \space undefined}}%
\hbox{\csname b@\@citeb\endcsname}}}{#1}}
\catcode`@=12

\def\citere#1{\mbox{Ref.~\cite{#1}}}

\newcommand{\mste}{m_{\tilde{t}_1}}
\newcommand{\mstz}{m_{\tilde{t}_2}}

\newcommand{\Xt}{X_t}

\newcommand{\msusy}{M_{\rm SUSY}}

\newcommand{\cHe}{\cH_1}
\newcommand{\cHz}{\cH_2}

\newcommand{\Pe}{\phi_1^0}
\newcommand{\Pz}{\phi_2^0}

\newcommand{\PePz}{\phi_1^0\phi_2^0}

\newcommand{\DRbar}{\ensuremath{\overline{\mathrm{DR}}}}

\def\order#1{\ensuremath{{\cal O}(#1)}}

\newcommand{\cH}{{\cal H}}

\newcommand{\twol}{two-loop}

\newcommand{\tc}{{\sc TwoCalc}}

\newcommand{\fh}{{\sc FeynHiggs}}
\newcommand{\sd}{{\sc SecDec}}

\newcommand{\MZ}{m_Z}

\newcommand{\MA}{m_{A^0}}
\newcommand{\mh}{m_h}

\newcommand{\Mh}{M_h}
\newcommand{\MH}{M_H}

\newcommand{\mhmax}{$\mh^{\rm max}$}

\newcommand{\mt}{m_t}

\newcommand{\Mgl}{m_{\tilde{g}}}

\newcommand{\tsf}{\theta\kern-.20em_{\tilde{f}}}
\newcommand{\tsfp}{\theta\kern-.20em_{\tilde{f}\prime}}
\newcommand{\tsq}{\theta\kern-.15em_{\tilde{q}}}

\newcommand{\VL}{\left( \begin{array}{c}}
\newcommand{\VR}{\end{array} \right)}
\newcommand{\ML}{\left( \begin{array}{cc}}
\newcommand{\MLd}{\left( \begin{array}{ccc}}
\newcommand{\MLv}{\left( \begin{array}{cccc}}
\newcommand{\MR}{\end{array} \right)}
\newcommand{\re}{\mathop{\mathrm{Re}}}

\newcommand{\tb}{\tan \beta}

\newcommand{\tev}{\,\, \mathrm{TeV}}
\newcommand{\gev}{\,\, \mathrm{GeV}}
\newcommand{\mev}{\,\, \mathrm{MeV}}

\newcommand{\BC}{\begin{center}}
\newcommand{\EC}{\end{center}}
\newcommand{\BE}{\begin{equation}}
\newcommand{\BEA}{\begin{eqnarray}}
\newcommand{\EEA}{\end{eqnarray}}
\newcommand{\non}{\nonumber}
\newcommand{\id}{{\rm 1\kern-.12em
\rule{0.3pt}{1.5ex}\raisebox{0.0ex}{\rule{0.1em}{0.3pt}}}}

\newcommand{\gsim}
{\;\raisebox{-.3em}{$\stackrel{\displaystyle >}{\sim}$}\;}

\def\als{\alpha_s}
\def\alt{\alpha_t}
\def\alb{\alpha_b}

\def\de{\delta}

\def\De{\Delta}

\def\hSi{\hat{\Sigma}}
\newcommand{\se}[1]{\Sigma_{#1}}

\newcommand{\ser}[1]{\hat{\Sigma}_{#1}}

\definecolor{Lightblue}{cmyk}{0.9,0.1,0.1,0.3}
\definecolor{dgelborange}{cmyk}{0.,0.3,0.5, 0.}
\definecolor{Lila}{rgb}{0.5,0.,1}

\graphicspath{{plots/}}
\newcommand{\MHexp}{125.6}
\title{Momentum Dependent Two-Loop Corrections to the Neutral Higgs Boson Masses in the MSSM}

\ShortTitle{MSSM Higgs-Boson masses at the two-loop level}

\author{\speaker{Sophia Borowka}\\
        Max Planck Institute for Physics, F\"ohringer Ring 6, 80805 Munich, Germany\\
        E-mail: \email{sborowka@mpp.mpg.de}}
	
\abstract{
The momentum dependent two-loop contributions of the order 
${\cal O} (\alt\als)$ to the masses in the Higgs-boson 
sector of the MSSM are computed.
Adopting the Feynman-diagrammatic approach and using a mixed
on-shell/\DRbar\ renormalization scheme, the new corrections can directly 
be matched onto the higher-order corrections included in the 
code \fh. Two-loop diagrams involving several mass scales 
are evaluated with the program \sd. 
The combination of the new momentum dependent two-loop contribution with
the existing one- and two-loop corrections  
leads to an improved prediction of the light
MSSM Higgs-boson mass with reduced theoretical
uncertainty. 
The resulting shifts in the lightest Higgs-boson mass 
$\Mh$ can extend up
to the level of the current experimental uncertainty of about 500\,MeV in 
the scenario considered in these proceedings.
}

\FullConference{Loops and Legs in Quantum Field Theory - LL 2014,\\
		 27 April - 2 May 2014 \\
		 Weimar, Germany }

\begin{document}

\section{Introduction}
\label{sec:intro}
The ATLAS and CMS experiments at CERN have discovered a
new boson with a mass around 
$\MHexp \gev$~\cite{ATLASdiscovery,CMSdiscovery}. 
Despite its seemingly  Standard Model-like behavior within the present 
experimental uncertainties, the newly discovered 
particle can also be interpreted as the Higgs-boson of extended models.
The Higgs-boson sector of the Minimal Supersymmetric
Standard Model (MSSM)~\cite{mssm} with two scalar doublets
accommodates five physical Higgs-bosons,  
the light and heavy $CP$-even bosons $h^0$
and $H^0$, the $CP$-odd boson $A^0$, and the charged Higgs-bosons $H^\pm$. 
The light $CP$-even Higgs-boson $h^0$ can be identified with the newly 
discovered scalar particle. Scenarios where the latter is associated with 
the heavy $CP$-even Higgs-boson $H^0$ is not considered in these proceedings. 
In the MSSM, the mass of $h^0$, $\Mh$, can directly be predicted from
the other parameters of the model. The accuracy of this prediction 
should at least match the one of the experimentally measured mass value for the 
new boson.

The status of higher-order corrections to the masses and mixing angles
in the neutral Higgs-boson sector of the MSSM with real parameters is quite advanced. 
The complete one-loop
result within the MSSM is known~\cite{ERZ,mhiggsf1lA,mhiggsf1lB,mhiggsf1lC}.
The dominant one-loop contributions are the ones of order  $\alt$  originating from
top and stop loops ($\alt \equiv y_t^2 / (4 \pi)$ and $y_t$ being the
top-quark Yukawa coupling). The range of available two-loop corrections
meanwhile also covers most of the contributions which are believed to 
be significant~\cite{mhiggsletter,mhiggslong,mhiggslle,mhiggsFD2,
bse,mhiggsEP0,mhiggsEP1,mhiggsEP1b,mhiggsEP2,mhiggsEP3,
mhiggsEP3b,mhiggsEP4,mhiggsEP4b,mhiggsRG1,mhiggsRG1a}.
In particular, the \order{\alt\als} contributions to the self-energies -- evaluated in the
Feynman-diagrammatic (FD) as well as in the effective potential (EP)
approach -- as well as the \order{\alt^2}, \order{\alb\als}, 
\order{\alt\alb} and \order{\alb^2} contributions  -- evaluated in the EP
approach -- 
are known for vanishing external momenta.  
The obtained results are publicly available in the code 
\fh~\cite{feynhiggs,mhiggslong,mhiggsAEC,mhcMSSMlong,Mh-logresum}.

An evaluation of the momentum dependence at the two-loop level in a calculation 
employing the \DRbar\ scheme was presented in \citere{mhiggs2lp2}.
A (nearly) full two-loop EP calculation,  
including even the leading three-loop corrections, has also been
published~\cite{mhiggsEP5}. However, within the EP method 
all contributions are evaluated at zero external momentum for the corresponding self-energies, 
in contrast to the FD method, which in principle allows non-vanishing
external momentum. Further, the calculation presented in Ref.~\cite{mhiggsEP5} 
is not publicly available as a computer code 
for Higgs-boson mass calculations. 
Subsequently, another leading three-loop
calculation of \order{\alt\als^2} has been performed~\cite{mhiggsFD3l},
using assumptions on the various {\small SUSY} mass hierarchies,
resulting in the code {\sc H3m} (which
adds the three-loop corrections to the \fh~result).
Most recently, a combination of the full one-loop result, supplemented
with leading and sub-leading two-loop corrections evaluated in the
Feynman-diagrammatic/effective potential approach and a resummation of the
leading and sub-leading logarithmic contributions from the scalar-top
sector has been published~\cite{Mh-logresum} and included in the latest version of
the code~\fh~\cite{feynhiggs,mhiggslong,mhiggsAEC,mhcMSSMlong,Mh-logresum}.

In these proceedings, the calculation of mass shifts resulting from 
the inclusion of the leading momentum-dependent 
\order{\alt\als} corrections to the neutral $CP$-even 
Higgs-boson masses is described for one representative scenario. 
Further scenarios and more details are found in Ref.~\cite{Borowka:2014wla}.
\section{Outline of the calculation}
\label{sec:calculation}
The MSSM requires two doublets ${\cal H}_1$ and ${\cal H}_2$ of complex scalar fields 
which read 
\begin{align}
\cHe = \VL \cHe^0 \\[0.5ex] \cHe^- \VR \; = \; \VL v_1 
      + \frac{1}{\sqrt2}(\phi_1^0 - i\chi_1^0) \\[0.5ex] -\phi_1^- \VR\,,\hspace{3pt} 
\cHz = \VL \cHz^+ \\[0.5ex] \cHz^0 \VR \; = \; \VL \phi_2^+ \\[0.5ex] 
        v_2 + \frac{1}{\sqrt2}(\phi_2^0 + i\chi_2^0) \VR\,.
\label{higgsfeldunrot}
\end{align}
 The vacuum expectation values $v_1$ and $v_2$ define the angle $\tan\beta=v_2/v_1$.
At tree level, the mass matrix of the neutral CP-even Higgs-bosons in the $(\phi_1^0,\phi_2^0)$ basis can be written as 
\begin{align}
\label{eq:nondiag}
M_{\text{Higgs}}^{2,\text{tree}}=\left( \begin{matrix} \MA^2\text{sin}^2\, \beta + \MZ^2\text{cos}^2\, \beta & 
-(\MA^2 + \MZ^2)\,\text{sin}\, \beta \text{cos}\, \beta \\ 
-(\MA^2 + \MZ^2)\,\text{sin}\, \beta \text{cos}\, \beta & 
\MA^2\text{cos}^2\, \beta + \MZ^2\text{sin}^2\, \beta \end{matrix} \right)\;,
\end{align}
where $\MA$ is the mass of the CP-odd neutral Higgs-boson $A^0$. 
The rotation to the basis formed by the mass eigenstates $H^0,h^0$ is given by 
\begin{align}
\label{eq:physbasis}
 \left( \begin{matrix} H^0 \\ h^0 \end{matrix} \right) =  \left( \begin{matrix} \textrm{cos} \,\alpha  & 
 \textrm{sin} \,\alpha \\ 
-\textrm{sin} \,\alpha & 
\textrm{cos} \, \alpha \end{matrix} \right) \left( \begin{matrix} \phi_1^0 \\ 
\phi_2^0 \end{matrix} \right)\;.
\end{align}
\subsection{Computational set-up}
\label{subsec:howselfenergies}
The higher-order corrected $CP$-even Higgs-boson masses in the
MSSM are obtained  from the corresponding propagators
dressed by their self-energies. 
The inverse propagator matrix in the $(\Pe, \Pz)$ basis is given 
by
\begin{align}
\label{eq:prop}
(\Delta_{\text{Higgs}})^{-1} = -\text{i}
\left( \begin{matrix} 
p^2 - m_{\phi_1}^2 + \hat{\Sigma}_{\phi_1}(p^2) & -m_{\phi_1\phi_2}^2 +\hat{\Sigma}_{\phi_1\phi_2}(p^2)\\ 
-m_{\phi_1\phi_2}^2 +\hat{\Sigma}_{\phi_1\phi_2}(p^2) & p^2 - m_{\phi_2}^2 + \hat{\Sigma}_{\phi_2}(p^2) 
\end{matrix} \right) \text{ ,}
\end{align}
where the $\hat{\Sigma}(p^2)$ denote the renormalized Higgs-boson 
self-energies, $p$ being the external momentum.

\begin{figure}[htb!]
\begin{center}
\subfigure[]{\raisebox{0pt}{\includegraphics[width=0.2\textwidth]{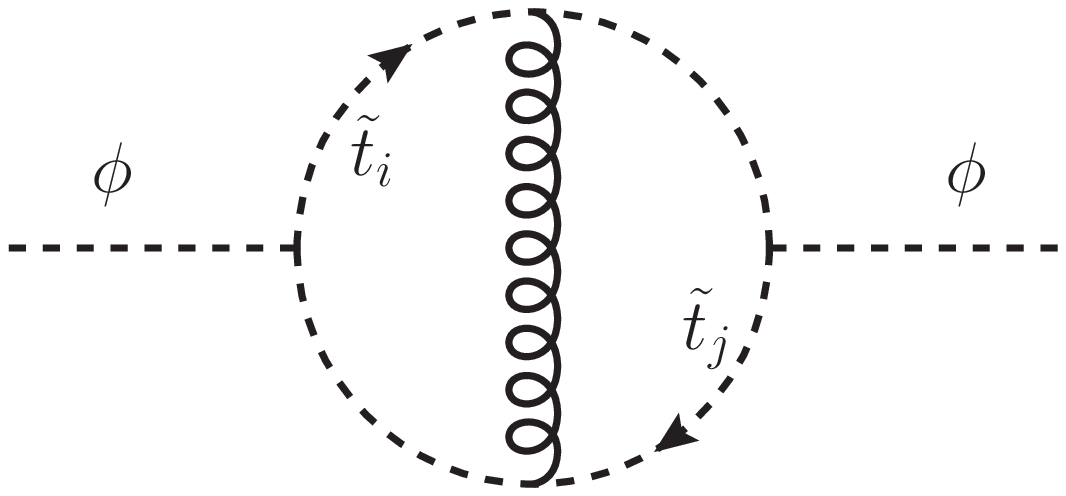}} }
\subfigure[]{\raisebox{1pt}{\includegraphics[width=0.2\textwidth]{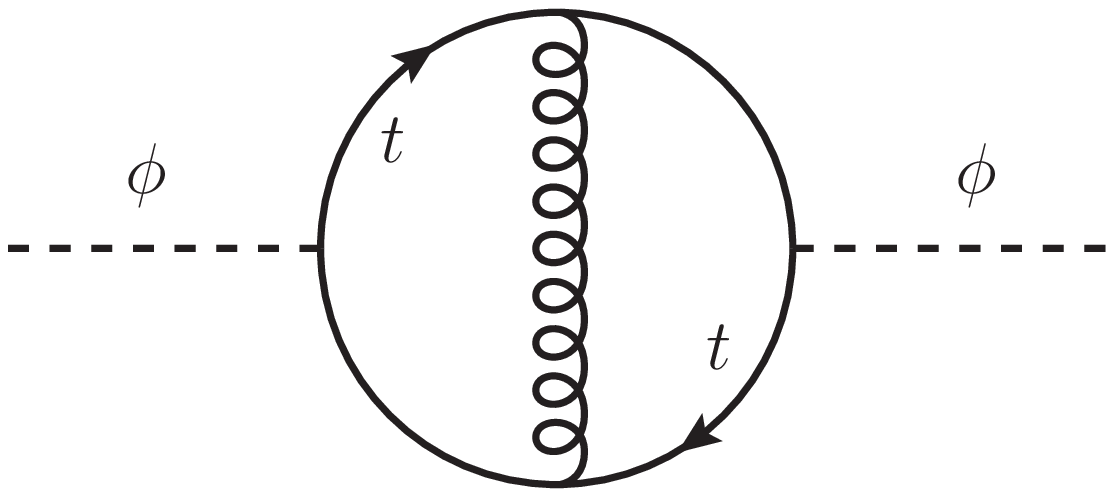}} }
\subfigure[]{\raisebox{1pt}{\includegraphics[width=0.2\textwidth]{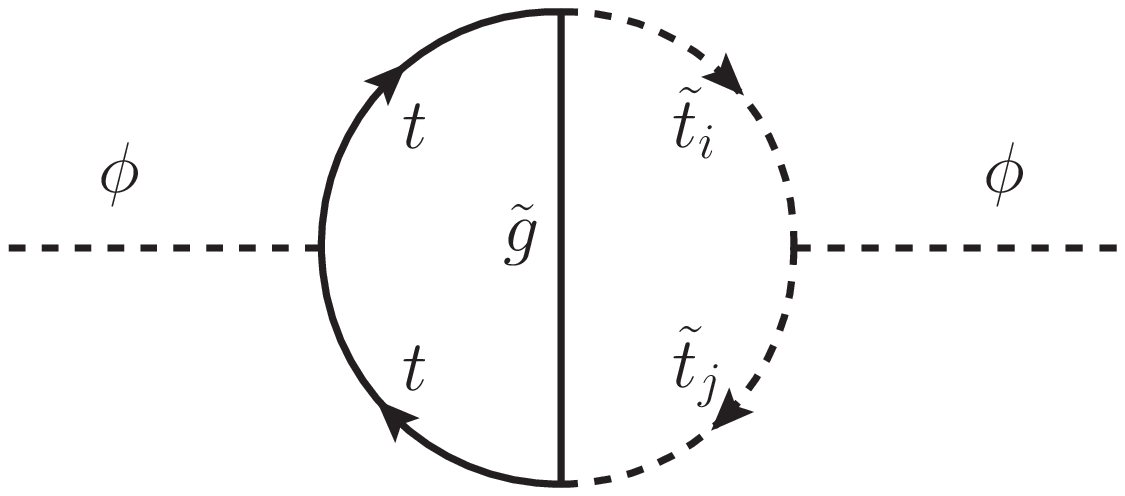}} }
\subfigure[]{\raisebox{-1pt}{\includegraphics[width=0.2\textwidth]{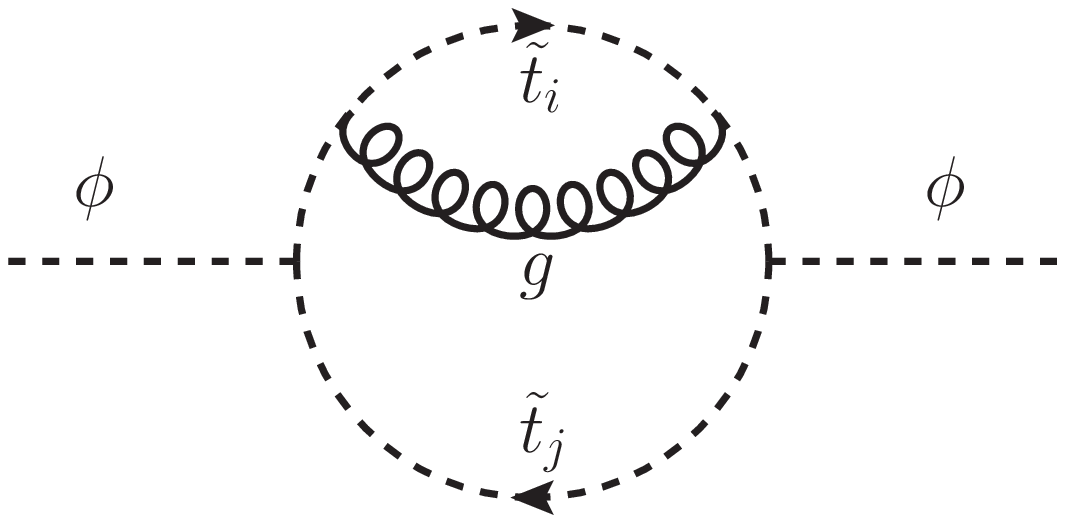}} }\\
\hspace{-10pt}\subfigure[]{\raisebox{0pt}{\includegraphics[width=0.2\textwidth]{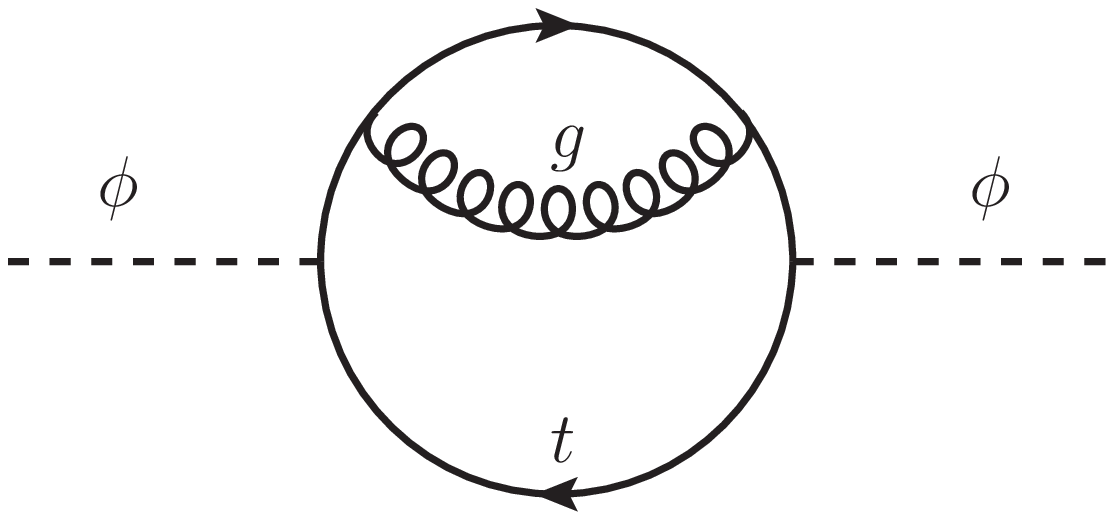}} }
\subfigure[]{\raisebox{-1pt}{\includegraphics[width=0.2\textwidth]{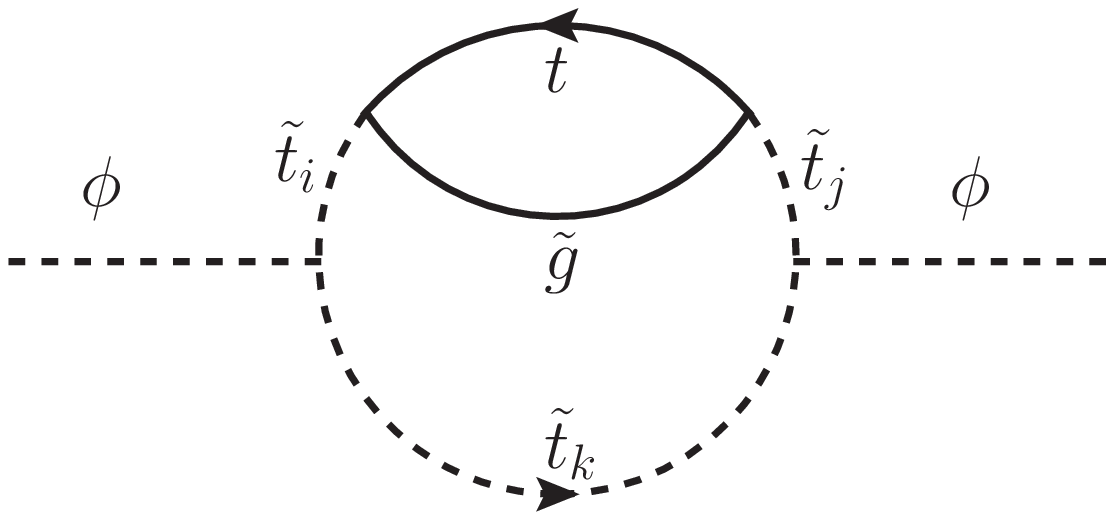}} }
\subfigure[]{\raisebox{0pt}{\includegraphics[width=0.2\textwidth]{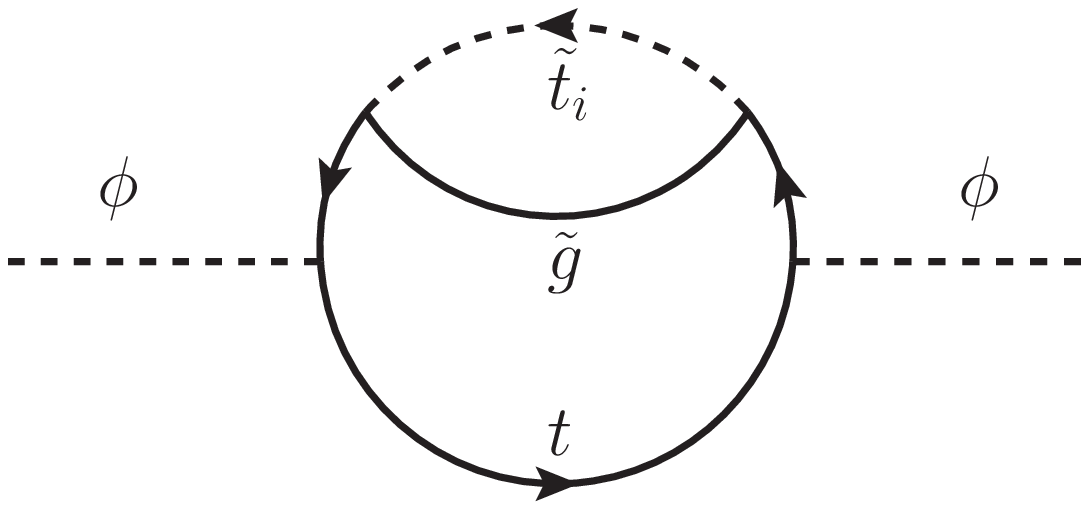}} }
\hspace{9pt}
\subfigure[]{\raisebox{0pt}{\includegraphics[width=0.15\textwidth]{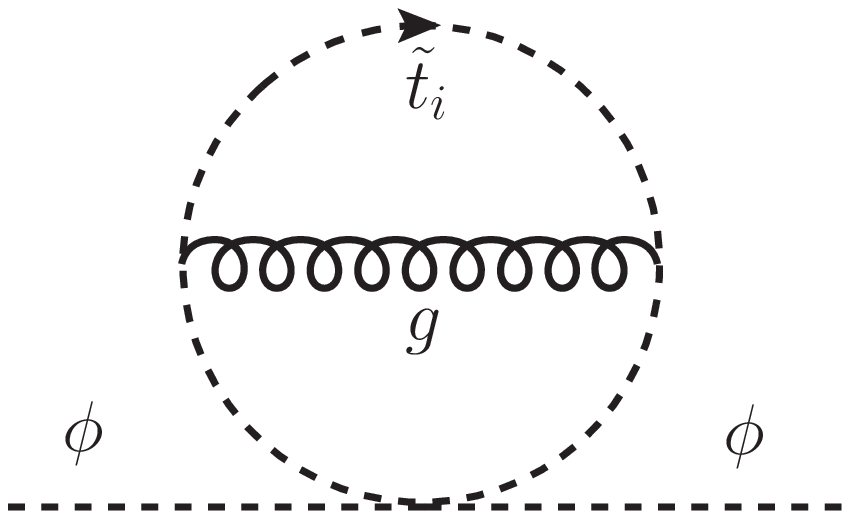}} }
\end{center} \vspace{-10pt}
\caption{Examples of  \twol\ diagrams enetring the Higgs-boson self-energies
($\phi = h^0, H^0, A^0$). }
\label{fig:fd_hHA}
\end{figure}

The calculation is performed in the Feynman-diagrammatic (FD) approach. To
obtain expressions for the unrenormalized self-energies at \order{\alt\als}, the 
evaluation of genuine two-loop diagrams
and one-loop graphs with counter-term insertions is required. 
Example diagrams for the neutral Higgs-boson self-energies are shown
in Fig.~\ref{fig:fd_hHA}.
For the counter-term insertions, 
one-loop diagrams with external top quarks/squarks have 
to be evaluated. In addition, two-loop tadpole 
diagrams enter the two-loop counter terms.  
The complete set of contributing Feynman diagrams 
has been generated with the
program {\sc FeynArts}~\cite{feynarts} (using the model file including
counter terms from \citere{mssmct}).
A tensor reduction and evaluation of traces was performed with 
the programs~{\sc FormCalc}~\cite{formcalc} and 
\tc~\cite{twocalc},
yielding algebraic expressions in terms of 
the scalar one- and two-point one-loop functions,  
massive two-loop vacuum functions~\cite{Davydychev:1992mt}, 
and two-loop integrals which depend on the external momentum.
The latter have been evaluated with the program 
\sd~\cite{Borowka:2012yc,Borowka:2013cma}. 
%
%
\subsection{Computation of mass shifts}
\label{subsec:howmassshifts}
The calculation of the self-energies is performed in the $(\phi_1^0,\phi_2^0)$ basis. 
To be consistent with the higher-order contributions to the 
Higgs-boson masses incorporated in the  program \fh, 
the renormalized self-energies in the $(\phi_1^0,\phi_2^0)$ basis are
rotated into the physical $(h^0,H^0)$ basis, 
\begin{subequations}
\BEA
\ser{H^0H^0}^{(2)}&=& 
\cos^2\!\alpha \,\ser{\Pe\Pe}^{(2)} + 
\sin^2\!\alpha \,\ser{\Pz\Pz}^{(2)} + 
\sin (2 \alpha) \, \ser{\PePz}^{(2)} \text{ ,}\\
\ser{h^0h^0}^{(2)} &=& 
\sin^2\!\alpha \,\ser{\Pe\Pe}^{(2)} + 
\cos^2\!\alpha \,\ser{\Pz\Pz}^{(2)} - 
\sin (2 \alpha) \,   \ser{\PePz}^{(2)} \text{ ,} \\
\ser{h^0H^0}^{(2)} &=& 
\sin \alpha \cos\alpha \,(\ser{\Pz\Pz}^{(2)} - \ser{\Pe\Pe}^{(2)}) + 
\cos (2 \alpha) \,   \,\ser{\PePz}^{(2)} \text{ ,}
\EEA
\label{eq:transformationphi12tohH}%
\end{subequations}%
where the tree-level 
propagator matrix is diagonal and $\alpha$ the tree-level mixing angle, 
see Eqs.~(\ref{eq:nondiag})-(\ref{eq:physbasis}). 
The resulting new contributions to the neutral $CP$-even Higgs-boson 
self-energies, containing all momentum-dependent and additional constant 
terms, are assigned to the differences
\begin{equation}
\De\ser{ab}^{(2)}(p^2) = \ser{ab}^{(2)}(p^2) - \tilde\Sigma_{ab}^{(2)}(0)\,,
\qquad
ab = \{H^0H^0,h^0H^0,h^0h^0\}\,.
\label{eq:DeltaSE}
\end{equation}
Note the tilde (not hat) on $\tilde\Sigma^{(2)}(0)$, which signifies that 
not only the self-energies are evaluated at zero external momentum but
also the corresponding counter terms,
following Refs.~\cite{Heinemeyer:1998jw,Heinemeyer:1998kz,
Heinemeyer:1998np}.
A finite shift $\De\hat{\Sigma}^{(2)}  (0)$
therefore remains in the limit $p^2\to 0$ 
due to $\de m_{A^0}^{2(2)} = \re\se{A^0A^0}^{(2)}(m_{A^0}^2)$ being computed 
at $p^2=m_{A^0}^2$ 
in $\hat\Sigma^{(2)}$, but at $p^2=0$ in $\tilde\Sigma^{(2)}$.

Several checks have been performed on the calculation.
Subtracting the finite shift of $\de m_{A^0}^{2(2)}$, the finite 
shift $\De\ser{ab}^{(2)}(0)$ in Eq.~(\ref{eq:DeltaSE}) must cancel 
in the limit of vanishing external momentum. 
This could be confirmed numerically. 
Moreover, agreement with previous calculations performed 
in the zero momentum limit \cite{Heinemeyer:1998jw,Heinemeyer:1998np} 
was found analytically. 
All integrals which were deduced analytically from known 
expressions~\cite{Davydychev:1992mt,Berends:1994sa} 
were checked with \sd.
For more details about the calculational set-up the reader is 
referred to \cite{Borowka:2014wla,Borowka:2013uea}.
\medskip

According to Eq.~(\ref{eq:prop}), 
the $CP$-even Higgs-boson masses are determined from the
poles of the $h^0$-$H^0$-propagator matrix. 
This is equivalent to solving the equation
\begin{equation}
\left[p^2 - m_{h^0}^2 + \hSi_{h^0h^0}(p^2) \right]
\left[p^2 - m_{H^0}^2 + \hSi_{H^0H^0}(p^2) \right] -
\left[\hSi_{h^0H^0}(p^2)\right]^2 = 0\,~,
\label{eq:proppole}
\end{equation}
yielding the loop-corrected pole masses, $\Mh$ and $\MH$.
%

\section{Numerical results}
\label{sec:numericalresults}

\begin{figure}[ht!]
\centering
\includegraphics[width=0.49\textwidth]{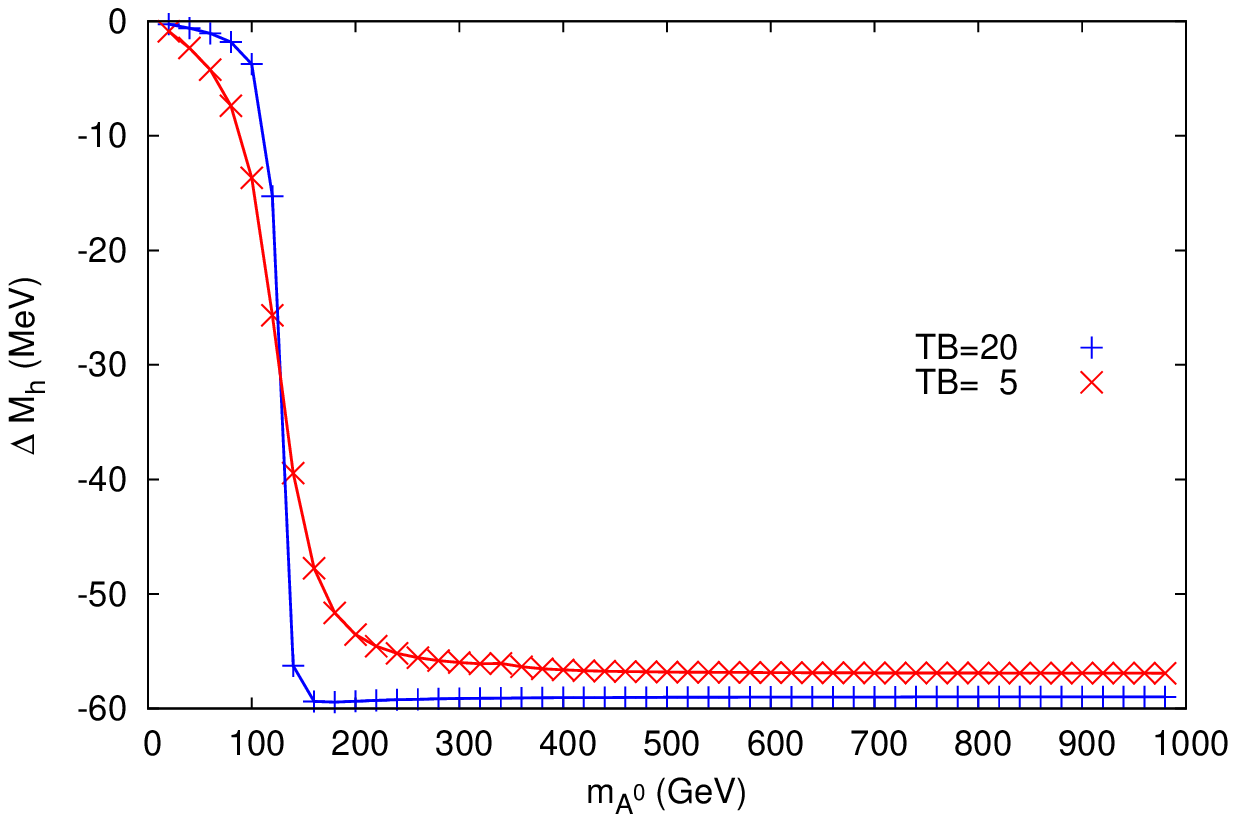}
\includegraphics[width=0.49\textwidth]{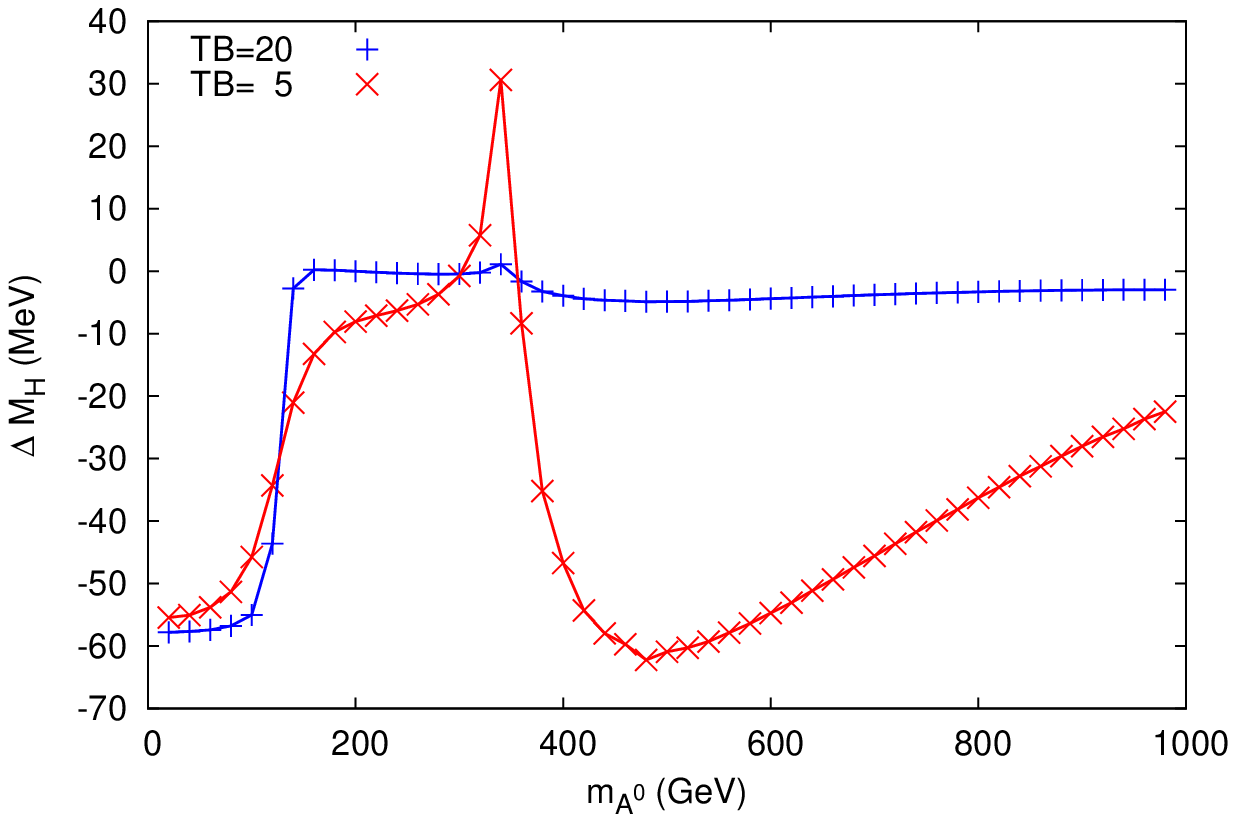}
\caption{Variation of the mass shifts $\Delta\Mh,\Delta\MH$ with the
  $A^0$-boson mass $\MA$ 
for $\tb=5$ (blue) and $\tb = 20$ (red). The  peak in $\Delta\MH$
originates from a threshold at $2\,\mt$.} 
\label{fig:shiftswithma}
\end{figure}

%
%
The following parameter values are adopted for the numerical studies shown below
\begin{align}
\mt &= 173.2\gev,\; \msusy=1\tev,\; \Xt =2\,\msusy\, ,\; \mu = 200\gev~,\non \\
\Mgl &= 1500\gev, \;
\mste = 826.8\gev,\;  \mstz= 1173.2\gev\, .
\end{align}
They are oriented at the \mhmax\ scenario described 
in~\citere{Carena:2013qia}. 
Results for other scenarios and more details can be found in Ref.~\cite{Borowka:2014wla}.
In Fig.~\ref{fig:shiftswithma}, $\De\Mh$ (left plot) and $\De\MH$ (right plot) 
are shown as a function of
$\MA$ for $\tb = 5$ (blue) and $\tb = 20$ (red). 
In the \mhmax\ scenario for $\MA \gsim 200 \gev$, the additional shift 
$\Delta\Mh \sim - 60 \mev$ amounts to the size of the anticipated experimental
precision at a linear collider. The contribution to the heavy $CP$-even
Higgs-boson mass can 
reach $-60 \mev$ for very small or intermediate values of $\MA$,
whereas for $\MA \gsim 500 \gev$ a decreasing correction to $\MH$ 
can be observed.
The peak in $\De\MH$ for $\tb = 5$ originates from a threshold at $2\,\mt$.

\begin{figure}[htb!]
\centering
\includegraphics[width=0.49\textwidth]{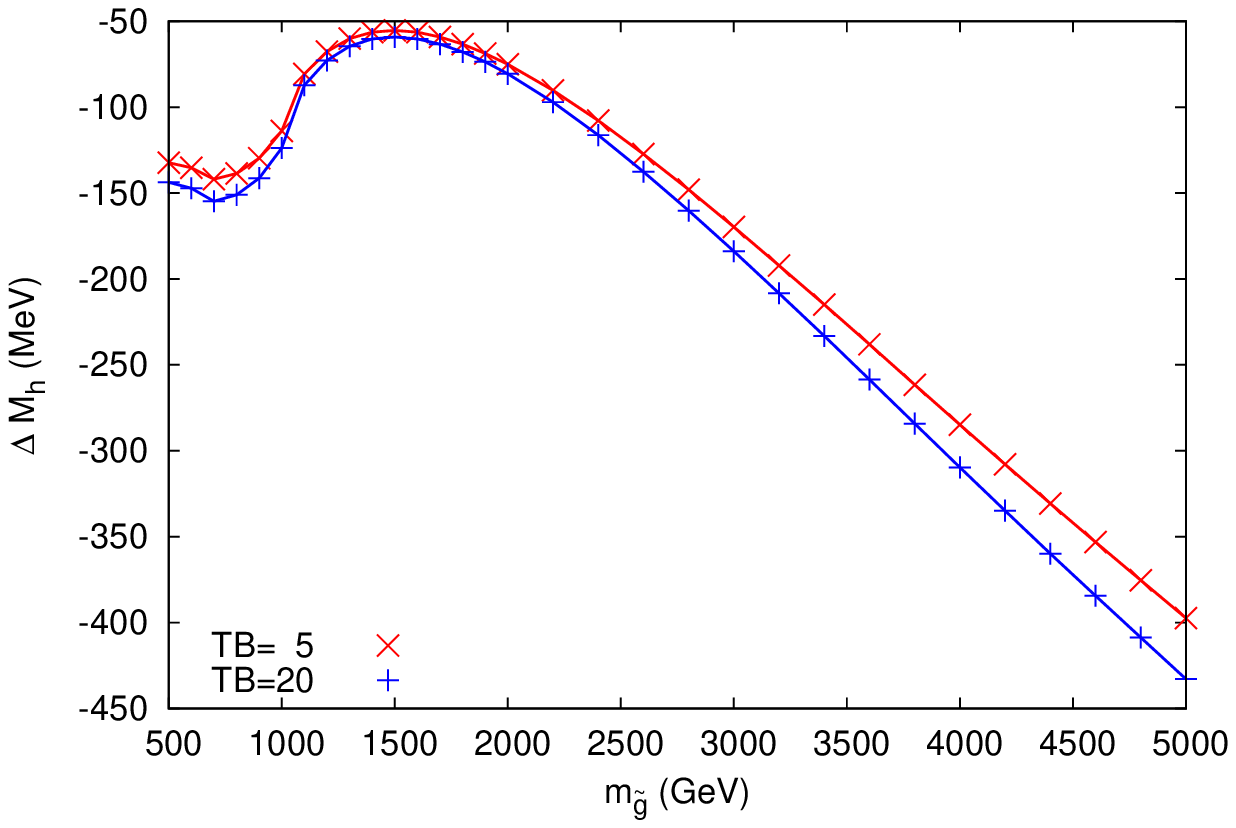}
\includegraphics[width=0.49\textwidth]{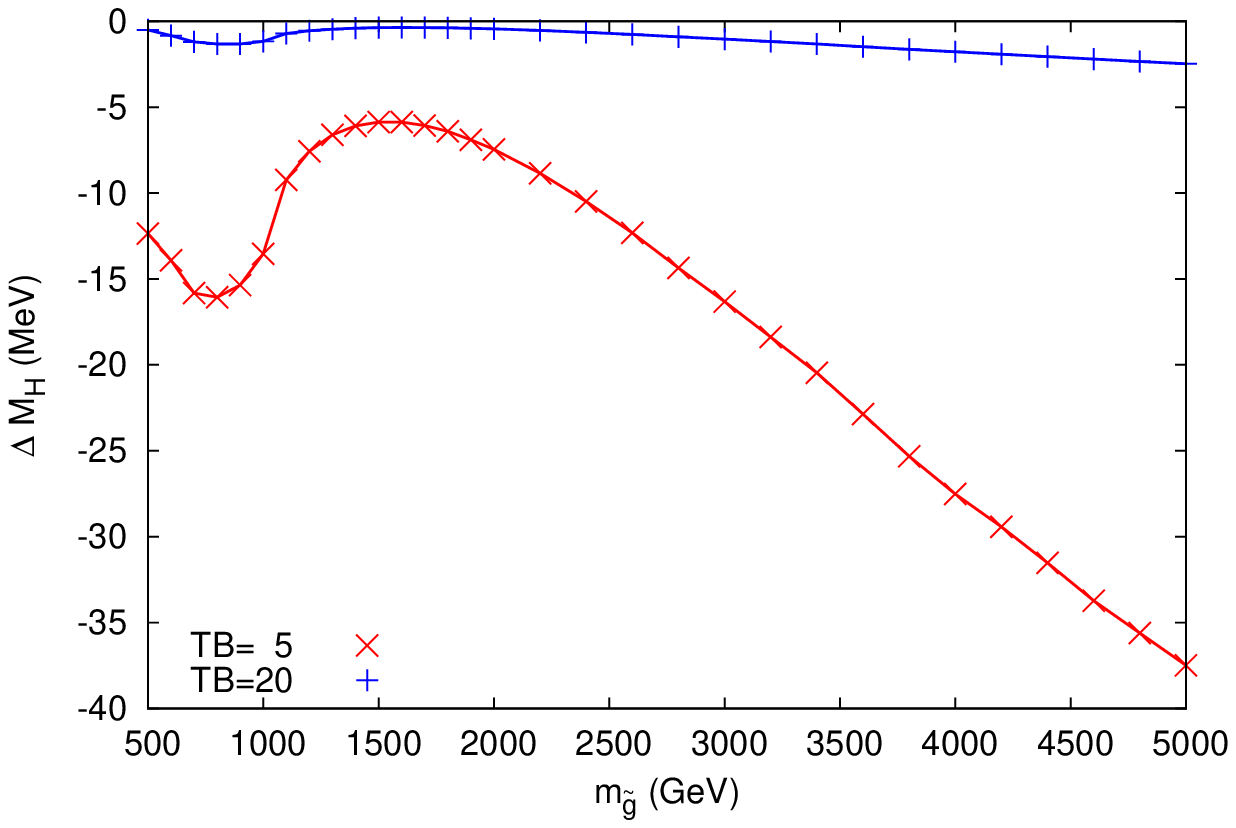}
\caption{Variation of the mass shifts $\Delta\Mh,\Delta\MH$ with the
  gluino mass for two different  values of
  $\tb=5,20$ and $\MA = 250 \gev$.
}
\label{fig:variationmgluino}
\end{figure} 

\medskip
Furthermore, the dependence of $\Mh$ and $\MH$ on the gluino mass $\Mgl$ is 
analyzed in the scenario described above. The results are shown in
Fig.~\ref{fig:variationmgluino} for $\De\Mh$ (left plot) and $\De\MH$
(right plot) for $\MA = 250 \gev$ 
with the same color coding as in Fig.~\ref{fig:shiftswithma}.
In the case of $\Mh$ one can observe that the effects are smallest
for $\Mgl\sim 1.5$\,TeV.
More sizable shifts occur for larger gluino masses,   
by more than $-400 \mev$ for $\Mgl \gsim 4 \tev$, 
reaching thus the level of
the current experimental accuracy in the Higgs-boson mass 
determination.
The corrections to $\MH$ 
do not exceed $-50 \mev$ in the considered $\Mgl$ range.


\section{Conclusion}
\label{sec:conclusion}
Results for the leading momentum-dependent
\order{\alt\als} contributions to the masses of the neutral 
$CP$-even Higgs-bosons in the MSSM have been presented. 
They were obtained by calculating
the corresponding contributions to the dressed Higgs-boson propagators 
in the Feynman-diagrammatic approach, using a mixed
on-shell/\DRbar\  renormalization scheme. 

\smallskip
The effect of the new momentum-dependent
two-loop corrections on the predictions for
the $CP$-even Higgs-boson masses was investigated.
The numerical analysis displayed a strong dependence of the light
$CP$-even Higgs-boson mass on the value of the
gluino mass. For values of $\Mgl \sim 1.5 \tev$ corrections to
$\Mh$ of about $ -50 \mev$ are found, while  
for very large gluino masses, $\Mgl \gtrsim 4 \tev$, 
the corrections can amount to the 
level of the current experimental accuracy, i.e. about $500 \mev$ at the LHC. 
\smallskip
The effects are mostly below the current and future anticipated 
experimental accuracies for the heavy $CP$-even Higgs-boson mass. 
The new results of \order{\alt\als} have been incorporated into the program \fh.
\subsection*{Acknowledgements}
I would like to thank Thomas Hahn, Sven Heinemeyer, 
Gudrun Heinrich and Wolfgang Hollik for the fruitful collaboration, and 
Stefano di Vita for comparisons. 
Furthermore I wish to thank the organizers of Loops and Legs 2014 for 
the nice and interesting conference.


\begin{thebibliography}{99}


\bibitem{ATLASdiscovery} G.~Aad et al. [ATLAS Collaboration],
  {\em Phys.\ Lett.} {\bf B 716} (2012) 1
  [arXiv:1207.7214 [hep-ex]].

\bibitem{CMSdiscovery} S.~Chatrchyan et al. [CMS Collaboration],
  {\em Phys.\ Lett.} {\bf B 716} (2012) 30
  [arXiv:1207.7235 [hep-ex]].

\bibitem{mssm} H.~Nilles, 
               {\em Phys.\ Rept.} {\bf 110} (1984) 1; \\ 
               H.~Haber and G.~Kane, 
               {\em Phys.\ Rept.} {\bf 117} (1985) 75; \\  
               R.~Barbieri, 
               {\em Riv.\ Nuovo Cim.} {\bf 11} (1988) 1. 




%
%


%



%

\bibitem{ERZ} J.~Ellis, G.~Ridolfi and F.~Zwirner,
              {\em Phys.\ Lett.} {\bf B 257} (1991) 83;\\
              Y.~Okada, M.~Yamaguchi and T.~Yanagida,
              {\em Prog.\ Theor.\ Phys. } {\bf 85} (1991) 1;\\
              H.~Haber and R.~Hempfling,
              {\em Phys.\ Rev.\ Lett.}  {\bf 66} (1991) 1815.
\bibitem{mhiggsf1lA} A.~Brignole,
                     {\em Phys. Lett.}\ {\bf B 281} (1992) 284.
\bibitem{mhiggsf1lB} P.~Chankowski, S.~Pokorski and J.~Rosiek,
                     {\em Phys. Lett.} {\bf B 286} (1992) 307;
                     {\em Nucl. Phys.} {\bf B 423} (1994) 437
                     [arXiv:hep-ph/9303309].
\bibitem{mhiggsf1lC} A.~Dabelstein,
                     {\em Nucl. Phys.} {\bf B 456} (1995) 25
                     [arXiv:hep-ph/9503443];
                     {\em Z. Phys.} {\bf C 67} (1995) 495
                     [arXiv:hep-ph/9409375].
\bibitem{mhiggsletter} S.~Heinemeyer, W.~Hollik and G.~Weiglein, 
                       {\em Phys. Rev.} {\bf D 58} (1998) 091701
                       [arXiv:hep-ph/9803277]; 
                       {\em Phys. Lett.} {\bf B 440} (1998) 296
                       [arXiv:hep-ph/9807423].

\bibitem{mhiggslong} S.~Heinemeyer, W.~Hollik and G.~Weiglein,
                     {\em Eur. Phys. J.} {\bf C 9} (1999) 343
                     [arXiv:hep-ph/9812472].

\bibitem{mhiggslle} S.~Heinemeyer, W.~Hollik and G.~Weiglein,
                    {\em Phys. Lett.} {\bf B 455} (1999) 179
                    [arXiv:hep-ph/9903404].

\bibitem{mhiggsFD2} S.~Heinemeyer, W.~Hollik, H.~Rzehak and G.~Weiglein,
                    {\em Eur. Phys. J.} {\bf C 39} (2005) 465
                    [arXiv:hep-ph/0411114].

\bibitem{bse} M.~Carena, H.~Haber, S.~Heinemeyer, W.~Hollik, C.~Wagner,
              and G.~Weiglein,
              {\em Nucl. Phys.} {\bf B 580} (2000) 29
              [arXiv:hep-ph/0001002].

\bibitem{mhiggsEP0}  R.~Zhang, 
                     {\em Phys.\ Lett. } {\bf B 447} (1999) 89
                     [arXiv:hep-ph/9808299];\\
                     J.~Espinosa and R.~Zhang, 
                     {\em JHEP} {\bf 0003} (2000) 026
                     [arXiv:hep-ph/9912236].

\bibitem{mhiggsEP1} G.~Degrassi, P.~Slavich and F.~Zwirner,
                    {\em Nucl. Phys.} {\bf B 611} (2001) 403
                    [arXiv:hep-ph/0105096].

\bibitem{mhiggsEP1b} R.~Hempfling and A.~Hoang, 
                     {\em Phys. Lett.} {\bf B 331} (1994) 99
                     [arXiv:hep-ph/9401219].

\bibitem{mhiggsEP2} A.~Brignole, G.~Degrassi, P.~Slavich and F.~Zwirner,
                    {\em Nucl. Phys.} {\bf B 631} (2002) 195
                    [arXiv:hep-ph/0112177].

\bibitem{mhiggsEP3} J.~Espinosa and R.~Zhang,
                    {\em Nucl. Phys.} {\bf B 586} (2000) 3
                    [arXiv:hep-ph/0003246]. 

\bibitem{mhiggsEP3b} J.~Espinosa and I.~Navarro,
                     {\em Nucl.\ Phys.} {\bf B 615} (2001) 82
                     [arXiv:hep-ph/0104047].

\bibitem{mhiggsEP4} A.~Brignole, G.~Degrassi, P.~Slavich and F.~Zwirner,
                    {\em Nucl. Phys.} {\bf B 643} (2002) 79
                    [arXiv:hep-ph/0206101].

\bibitem{mhiggsEP4b} G.~Degrassi, A.~Dedes and P.~Slavich,
                    {\em Nucl. Phys.} {\bf B 672} (2003) 144
                    [arXiv:hep-ph/0305127].

\bibitem{mhiggsRG1} M.~Carena, J.~Espinosa, M.~Quir\'os and C.~Wagner, 
                    {\em Phys. Lett.} {\bf B 355} (1995) 209
                    [arXiv:hep-ph/9504316];\\
                    M.~Carena, M.~Quir\'os and C.~Wagner, 
                    {\em Nucl. Phys.} {\bf B 461} (1996) 407
                    [arXiv:hep-ph/9508343].

\bibitem{mhiggsRG1a} J.~Casas, J.~Espinosa, M.~Quir\'os and A.~Riotto,
                     {\em Nucl. Phys.} {\bf B 436} (1995) 3,
                     [Erratum-ibid.\ {\bf B 439} (1995) 466]
                     [arXiv:hep-ph/9407389].

\bibitem{feynhiggs} S.~Heinemeyer, W.~Hollik and G.~Weiglein,
                    {\em Comput. Phys. Commun.} {\bf 124} (2000) 76
                    [arXiv:hep-ph/9812320];\\
                    T.~Hahn, S.~Heinemeyer, W.~Hollik, H.~Rzehak and
                    G.~Weiglein, 
                    {\em Comput.\ Phys.\ Commun.} {\bf 180} (2009) 1426; 
                    see: {\tt www.feynhiggs.de} .
\bibitem{mhiggs2lp2} S.~Martin,
                    {\em Phys. Rev.} {\bf D 71} (2005) 016012
                    [arXiv:hep-ph/0405022].

\bibitem{mhiggsEP5} S.~Martin, 
                    {\em Phys. Rev.} {\bf D 65} (2002) 116003
                    [arXiv:hep-ph/0111209];
                    {\em Phys. Rev.} {\bf D 66} (2002) 096001
                    [arXiv:hep-ph/0206136];
                    {\em Phys. Rev.} {\bf D 67} (2003) 095012
                    [arXiv:hep-ph/0211366];
                    {\em Phys. Rev.} {\bf D 68} (2003) 075002
                    [arXiv:hep-ph/0307101]; 
                    {\em Phys. Rev.} {\bf D 70} (2004) 016005
                    [arXiv:hep-ph/0312092];
                    {\em Phys. Rev.} {\bf D 71} (2005) 116004
                    [arXiv:hep-ph/0502168];
                    {\em Phys.\ Rev.} {\bf D 75} (2007) 055005
                    [arXiv:hep-ph/0701051];\\
                    S.~Martin and D.~Robertson,
                    {\em Comput.\ Phys.\ Commun.} {\bf 174} (2006) 133
                    [arXiv:hep-ph/0501132].

\bibitem{mhiggsFD3l} R.~Harlander, P.~Kant, L.~Mihaila and M.~Steinhauser,
                     {\em Phys.\ Rev.\ Lett.} {\bf 100} (2008) 191602
                     [{\em Phys.\ Rev.\ Lett.} {\bf 101} (2008) 039901]
                     [arXiv:0803.0672 [hep-ph]];
                     {\em JHEP} {\bf 1008} (2010) 104
                     [arXiv:1005.5709 [hep-ph]].

\bibitem{Mh-logresum} T.~Hahn, S.~Heinemeyer, W.~Hollik, H.~Rzehak and
                      G.~Weiglein, 
                      {\em Phys. Rev. Lett.} {\bf 112} (2014) 141801, 
                      arXiv:1312.4937 [hep-ph].

\bibitem{mhiggsAEC} G.~Degrassi, S.~Heinemeyer, W.~Hollik,
                    P.~Slavich and G.~Weiglein, 
                    {\em Eur. Phys. J.} {\bf C 28} (2003) 133
                    [arXiv:hep-ph/0212020].

\bibitem{mhcMSSMlong} M.~Frank, T.~Hahn, S.~Heinemeyer, W.~Hollik,  
                      H.~Rzehak and G.~Weiglein,
                      {\em JHEP} {\bf 0702} (2007) 047
                      [arXiv:hep-ph/0611326].

\bibitem{Borowka:2014wla} S.~Borowka, T.~Hahn, S.~Heinemeyer, G.~Heinrich 
			and W.~Hollik,
			{\em Eur. Phys. J.} {\bf C 74} (2014) 2994
			arXiv:1404.7074 [hep-ph].

\bibitem{feynarts} J.~K\"ublbeck, M.~B\"ohm and A.~Denner, 
                   {\em Comput. Phys. Commun.} {\bf 60} (1990) 165;\\
                   T.~Hahn, 
                   {\em Comput. Phys. Commun.} {\bf 140} (2001) 418
                   [arXiv:hep-ph/0012260];\\
                   T.~Hahn and C.~Schappacher, 
                   {\em Comput. Phys. Commun.} {\bf 143} (2002) 54
                   [arXiv:hep-ph/0105349].\\
                   The program and the user's guide 
                   are available via {\tt www.feynarts.de} .

\bibitem{mssmct} T.~Fritzsche, T.~Hahn, S.~Heinemeyer, F.~von der
  Pahlen, H.~Rzehak and C.~Schappacher, 
  {\em Comput. Phys. Commun.} {\bf 185} (2014) 1529, 
  arXiv:1309.1692 [hep-ph].

\bibitem{formcalc} T.~Hahn and M.~P\'erez-Victoria,
                   {\em Comput. Phys. Commun.} {\bf 118} (1999) 153
                   [arXiv:hep-ph/9807565].

\bibitem{twocalc} G.~Weiglein, R.~Scharf and M.~B\"ohm,
                  {\em Nucl. Phys.} {\bf B 416} (1994) 606
                  [arXiv:hep-ph/9310358];\\
G.~Weiglein, R.~Mertig, R.~Scharf and M.~B\"ohm, 
in {\it New Computing Techniques in Physics Research 2},
ed.~D.~Perret-Gallix (World Scientific, Singapore, 1992), p.~617.

\bibitem{Davydychev:1992mt}
	A.~I.~Davydychev and J.~B.~Tausk,
	{\em Nucl.\ Phys.} {\bf B 397} (1993) 123.


  
\bibitem{Borowka:2012yc}
  S.~Borowka, J.~Carter and G.~Heinrich,
  {\em Comput.\ Phys.\ Commun.} {\bf 184} (2013) 396
  [arXiv:1204.4152 [hep-ph]].

\bibitem{Borowka:2013cma}
  S.~Borowka and G.~Heinrich,
  Comput.\ Phys.\ Commun.\  {\bf 184} (2013) 2552
  [arXiv:1303.1157 [hep-ph]].
  
 \bibitem{Heinemeyer:1998jw}
  S.~Heinemeyer, W.~Hollik and G.~Weiglein,
  Phys.\ Rev.\ D {\bf 58} (1998) 091701
  [hep-ph/9803277].

\bibitem{Heinemeyer:1998kz}
  S.~Heinemeyer, W.~Hollik and G.~Weiglein,
  Phys.\ Lett.\ B {\bf 440} (1998) 296
  [hep-ph/9807423].

\bibitem{Heinemeyer:1998np}
  S.~Heinemeyer, W.~Hollik and G.~Weiglein,
  Eur.\ Phys.\ J.\ C {\bf 9} (1999) 343
  [hep-ph/9812472].

\bibitem{Berends:1994sa}
  F.~A.~Berends, A.~I.~Davydychev, V.~A.~Smirnov and J.~B.~Tausk,
  Nucl.\ Phys.\ B {\bf 439} (1995) 536
  [hep-ph/9410232].

\bibitem{Borowka:2013uea}
  S.~Borowka and G.~Heinrich,
  PoS RADCOR {\bf 2013} (2014) 009
  [arXiv:1311.6476 [hep-ph]].
  
\bibitem{Carena:2013qia}
  M.~Carena, S.~Heinemeyer, O.~St{\aa}l, C.~Wagner and G.~Weiglein,
  {\em Eur.\  Phys.\  J.} {\bf C 73} (2013) 2552
  [arXiv:1302.7033 [hep-ph]].













  



  
%



\end{thebibliography}
\end{document}